\documentclass[aps,prab,reprint]{revtex4-2}

\usepackage{amsmath,amssymb}
\usepackage{graphicx}

\begin{document}

\title{Quasi-Strong-Strong Beam--Beam Modeling of Bootstrapping Injection in FCC-ee}

\author{Takashi~Mori}
\affiliation{%
High Energy Accelerator Research Organization (KEK),\\
Tsukuba, Ibaraki 305--0801, Japan
}
\email{takashi.mori@kek.jp}

\author{Katsunobu~Oide}
\affiliation{UNIGE, Universit\'e de Gen\`{e}ve, Geneva, Switzerland}
\affiliation{CERN, European Organization for Nuclear Research, Geneva, Switzerland}

\author{Frank Zimmermann}
\affiliation{CERN, European Organization for Nuclear Research, Geneva, Switzerland}

\begin{abstract}
The FCC-ee is designed to operate with exceptionally strong beam--beam interactions, making continuous injection a critical and non-trivial aspect of its operation. During the injection process, an unavoidable charge imbalance between the two colliding beams leads to asymmetric beam--beam forces, potentially compromising transverse stability.

In this paper, we introduce a quasi-strong-strong (QSS) beam--beam scheme, implemented in the SAD simulation framework. The method preserves a self-consistent beam--beam lens by coupling paired weak--strong simulations, while avoiding the computational cost of full strong--strong tracking. The injection process is modeled as a gradual increase of the stored bunch population, allowing the isolated study of beam--beam--driven optics deformation under charge imbalance.

Using the QSS approach, we investigate the feasibility of bootstrapping injection in the Z, W, and H operating modes of FCC-ee. Stable injection paths up to the nominal bunch population are identified in the W and H modes. In contrast, in the explored parameter region, the Z mode exhibits saturation of the stored population below the nominal value.
\end{abstract}

\maketitle

\section{Introduction}
The Future Circular Collider in the electron--positron configuration (FCC-ee) is designed to deliver unprecedented luminosity over a wide range of center-of-mass energies~\cite{FCCeeCDR,FCCeeParameters}. High luminosity is achieved through large bunch populations, small vertical beta functions, and the use of crab-waist optics~\cite{CrabWaist}.\par

Under these conditions, continuous injection becomes an essential component of machine operation. During the injection phase, colliding beams inevitably experience charge imbalance, leading to asymmetric beam--beam forces and dynamically evolving optics.\par

Conventional weak--strong simulations neglect feedback from beam--beam--driven optics deformation, while full strong--strong simulations are often computationally impractical for systematic injection studies. For injection stability, the dominant effects arise from the nonlinear beam--beam lens acting on the stored beam optics.\par

In this paper we introduce a quasi-strong-strong (QSS) beam--beam simulation scheme, implemented in the SAD framework~\cite{SAD}, tailored to the study of bootstrapping injection under strong beam--beam interaction. The method preserves self-consistency of the beam--beam lens while remaining computationally efficient. An important feature of the present implementation is that simulations are performed using the full FCC-ee lattice, naturally incorporating lattice-dependent effects.\par

The present study focuses on operating regimes where strong beam--beam interaction and continuous injection are most critical, as encountered in the Z, W, and H modes of FCC-ee. Using this framework, we investigate injection feasibility in the Z, W, and H operating modes. While stable injection to the nominal bunch population is demonstrated in the W and H modes, the Z mode shows saturation below the nominal value within the explored parameter space.

\section{Beam--Beam Regime of FCC-ee}
The beam--beam regime foreseen for FCC-ee significantly exceeds that of previous circular electron--positron colliders~\cite{FCCeeCDR}. The exceptionally large beam--beam parameters imply strong nonlinear optics deformation and enhanced sensitivity to multi-dimensional dynamics, especially during injection when charge imbalance is unavoidable.

\section{Quasi-Strong-Strong Beam--Beam Scheme}
\subsection{Motivation}
The extremely large beam--beam parameters foreseen for FCC-ee necessitate a dedicated treatment of beam--beam interaction during injection.

\subsection{Concept of the Quasi-Strong-Strong Scheme}
The QSS scheme couples paired weak--strong simulations to update the beam--beam interaction in a partially self-consistent manner.

The two colliding beams are referred to as Beam-0 and Beam-1 throughout this paper.
Each beam is tracked independently through the full lattice.

At each interaction point $k$, the beam--beam interaction is computed using the beam moments evaluated immediately upstream of the collision:
\begin{equation}
  \mathcal{P}_1^{(k)} = \mathcal{M}\bigl[\mathcal{B}_0(s_k^-)\bigr],
\end{equation}
where $\mathcal{B}_0(s_k^-)$ denotes the phase-space moments of Beam-0 at the location just before interaction point $k$, obtained through lattice tracking.
The quantity $\mathcal{B}_0(s_k^-)$ is mapped to the beam--beam lens acting on Beam-1.
The same procedure is applied symmetrically for Beam-1.

This formulation incorporates the full lattice transport between interaction points and provides a partially self-consistent approximation to strong--strong beam--beam dynamics.
In the present implementation, the effective strong-beam parameters are constructed from the statistical moments of the opposing beam distribution.
These include the beam centroids, transverse beam sizes, and selected second-order moments that characterize the phase-space structure.

Specifically, the parameters supplied to the beam--beam lens are derived from the phase-space variables $(x, y, x', y', z, \delta)$,
together with their second-order moments, including both transverse and longitudinal components,
as well as transverse--longitudinal correlations such as $\langle xz \rangle$, $\langle x'z \rangle$, $\langle yz \rangle$, and $\langle y'z \rangle$.

This formulation enables the beam--beam interaction to reflect the evolving three-dimensional structure of the beam distribution,
including coupled transverse--longitudinal dynamics in both transverse planes.

In this formulation, the beam--beam kick is applied immediately to each beam at the interaction point.
However, the interaction is evaluated using the opposing beam distribution obtained from lattice tracking up to the collision point,
resulting in a sequential (asynchronous) update of the two beams.

This introduces a bidirectional coupling without requiring a fully simultaneous solution,
and naturally incorporates lattice-dependent transport effects between interaction points.
The coupling is therefore not fully self-consistent in a simultaneous sense,
but is constructed through iterative exchange of beam moments at each interaction point.

\subsection{Scope and Validity of the Approximation}
The QSS scheme targets beam--beam--driven optics deformation during injection. Effects such as detailed collective mode dynamics are beyond the present scope.
In particular, coherent beam--beam eigenmodes are not explicitly resolved within the present formulation.
In limiting cases, the QSS scheme reduces to conventional weak--strong tracking when the strong-beam parameters are fixed, and approaches strong--strong behavior when the update is fully self-consistent at each interaction point.
This provides a consistency check on the formulation.

The present implementation includes transverse--longitudinal correlations (e.g. $\langle xz \rangle$ and $\langle yz \rangle$),
allowing for a three-dimensional description of beam--beam-driven dynamics.
Further extensions to include energy--transverse correlations (e.g. $\langle x\delta \rangle$ and $\langle x'\delta \rangle$)
would provide a more complete treatment of dispersion effects, but are left for future work.

\subsection{Relation to Existing Beam--Beam Models}
The QSS approach occupies an intermediate position between conventional weak--strong and full strong--strong simulations.
The QSS scheme is thus not intended to replace full strong--strong simulations, but to complement them in the specific context of injection dynamics.
In contrast to conventional weak--strong simulations, where the strong beam is fixed,
the QSS scheme introduces a dynamical update of the beam--beam lens based on the opposing beam.
At the same time, it avoids solving fully self-consistent strong--strong dynamics,
which would require simultaneous tracking of both beams with full coupling.
The QSS approach therefore provides an intermediate description that retains essential
beam--beam-driven optics deformation while remaining computationally tractable.

\section{Charge Imbalance and Injection Model}
The injection process is modeled as a ramp-up of the stored bunch population. The normalized population ratio is defined as
\begin{equation}
 r = \frac{N}{N_{\mathrm{nom}}}.
\end{equation}
Injection is applied as discrete increments
\begin{equation}
 r \rightarrow r + \Delta r_{\mathrm{I}} .
\end{equation}
The choice of $\Delta r_{\mathrm{I}}$ and the injection interval effectively controls the rate at which charge imbalance is introduced into the system.
This rate determines whether the beam--beam interaction can adapt adiabatically to the evolving optics or instead drives instability.
The injection process can therefore be understood as a dynamical competition between controlled charge increase and beam--beam-induced particle loss.

\section{Results}
\subsection{Stable Injection Paths in the W and H Modes}
\begin{figure}
  \includegraphics[width=\linewidth]{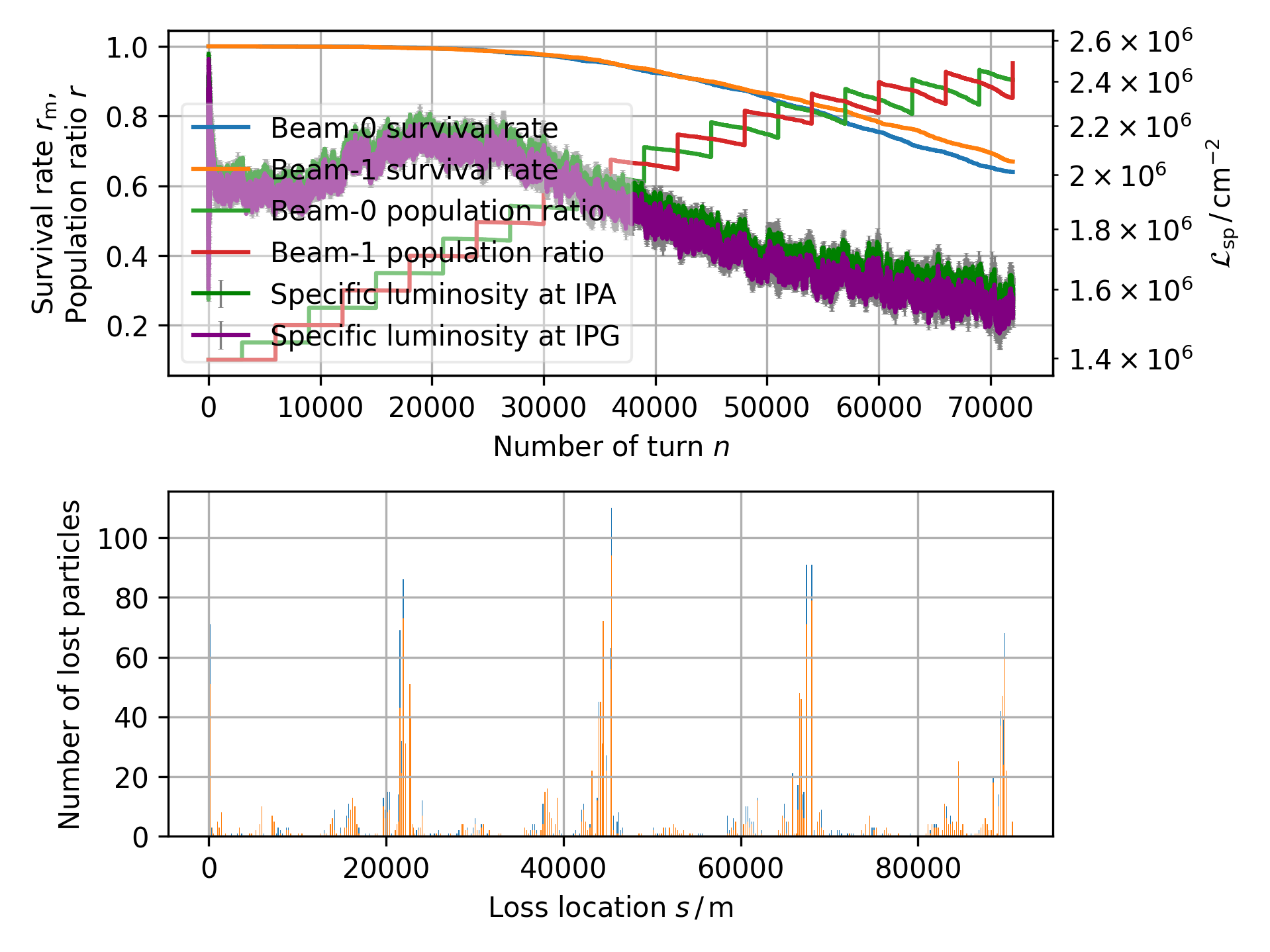}
  \caption{
    Evolution of the normalized bunch population, $r = N/N_{\mathrm{nom}}$, as a function of turn number $n$ during the injection process in the W mode, simulated using the quasi-strong-strong beam--beam scheme.
    Also shown are the macroparticle survival rate and the specific luminosity, included as auxiliary indicators of beam loss and collision performance.
    For the injection parameters shown, the stored beam population is stably ramped up to the nominal value without particle loss.
  }
  \label{fig:population:W mode}
\end{figure}

\begin{figure}
  \includegraphics[width=\linewidth]{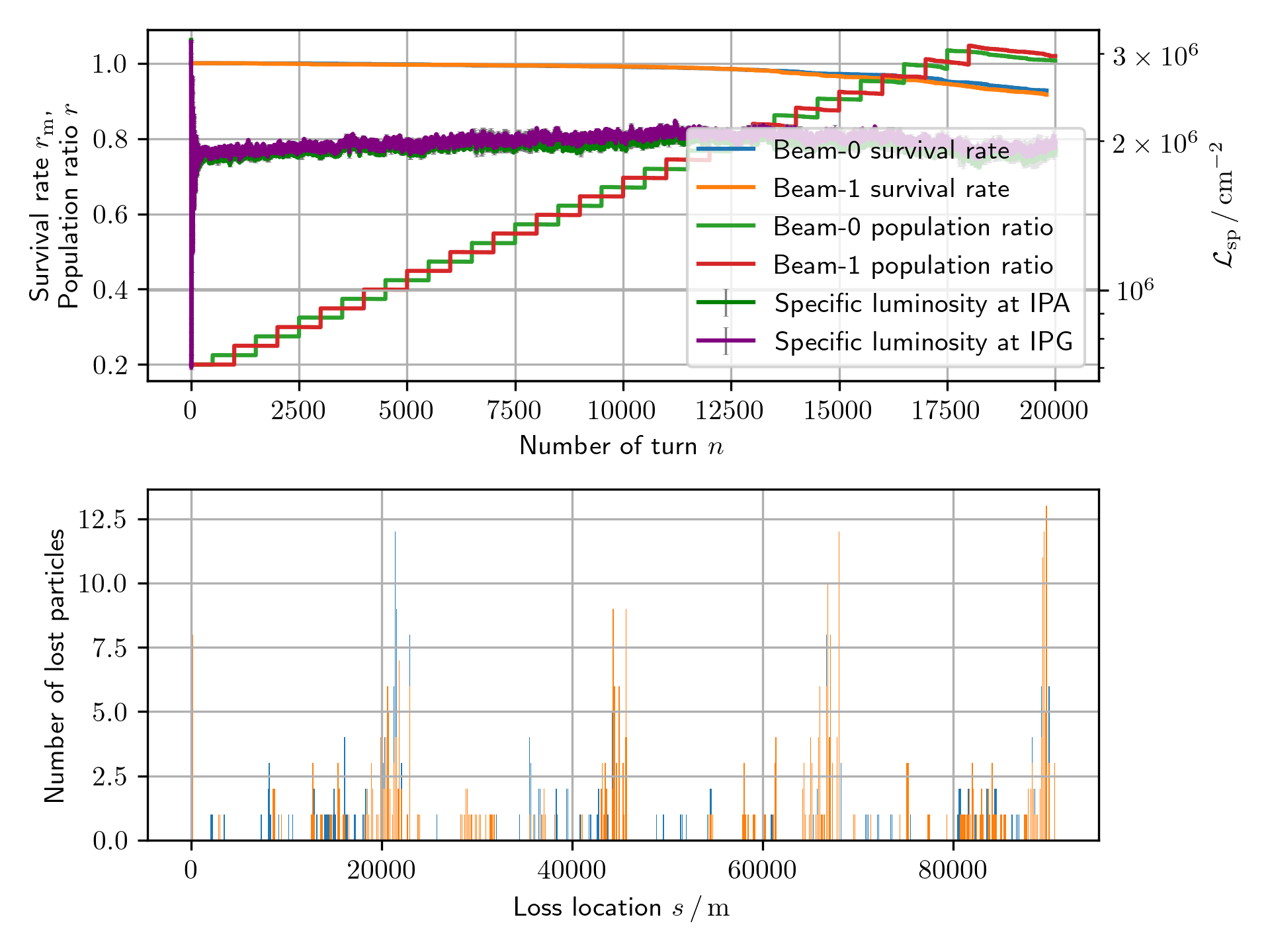}
  \caption{
    Same as Fig.~\ref{fig:population:W mode}, but for the H mode.
    The macroparticle survival rate and the specific luminosity evolve consistently with the population ramp-up, indicating stable injection up to the nominal bunch population under strong beam--beam interaction at the Higgs operating energy.
  }
  \label{fig:population:H mode}
\end{figure}

Representative examples of stable injection in the W and H modes are shown in Figs.~\ref{fig:population:W mode} and~\ref{fig:population:H mode}.
The macroparticle survival rate and the specific luminosity, shown together with the bunch population, provide complementary diagnostics indicating that beam loss and collision performance remain well controlled throughout the injection process on a turn-by-turn basis.
The specific luminosity is defined as the luminosity normalized to the bunch populations, thereby removing the trivial dependence on the total charge.
It reflects the beam geometry, bunch shape, and collision conditions at the interaction point, and is therefore used as a diagnostic of beam--beam quality during injection.

\subsection{Injection--Loss Balance in the Z Mode}
In the Z mode, injection saturates below the nominal population due to beam loss balancing the injected charge.
Coupled transverse--longitudinal dynamics associated with beam--beam interaction have been discussed in earlier studies of synchro--betatron effects~\cite{OhmiSynchroBeta}.
While the present operating conditions and modeling assumptions differ,
the observed emittance growth and x--z correlations indicate the development of coupled transverse--longitudinal dynamics.

In particular, the relatively weak radiation damping in the low-energy Z mode reduces the stabilization of longitudinal motion,
allowing such coupling to build up over many turns.
This leads to an effective increase in beam size and energy spread, which enhances particle loss through the nonlinear beam--beam interaction,
ultimately resulting in the observed injection--loss balance.
This behavior is consistent with an incoherent beam response to the evolving beam--beam interaction within the QSS framework.

\begin{figure}
  \includegraphics[width=\linewidth]{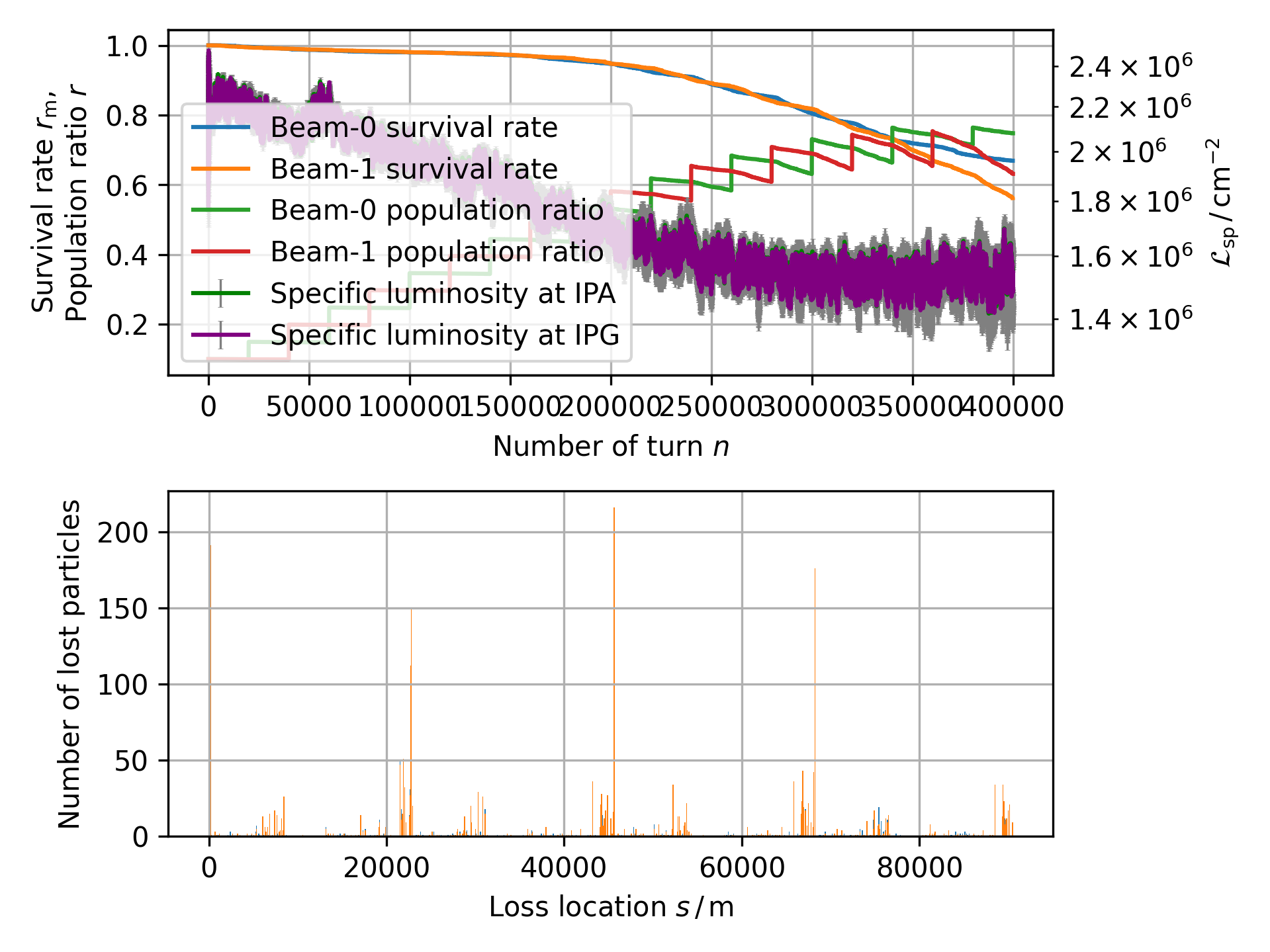}
  \caption{
    Evolution of the normalized bunch population, $r = N/N_{\mathrm{nom}}$, as a function of turn number $n$ in the Z mode.
    The macroparticle survival rate and the specific luminosity are shown simultaneously, illustrating the emergence of an injection--loss balance within the explored parameter range, where beam loss compensates the charge increase provided by injection.
  }
\end{figure}

\begin{figure}
  \centering
  \includegraphics[width=\linewidth]{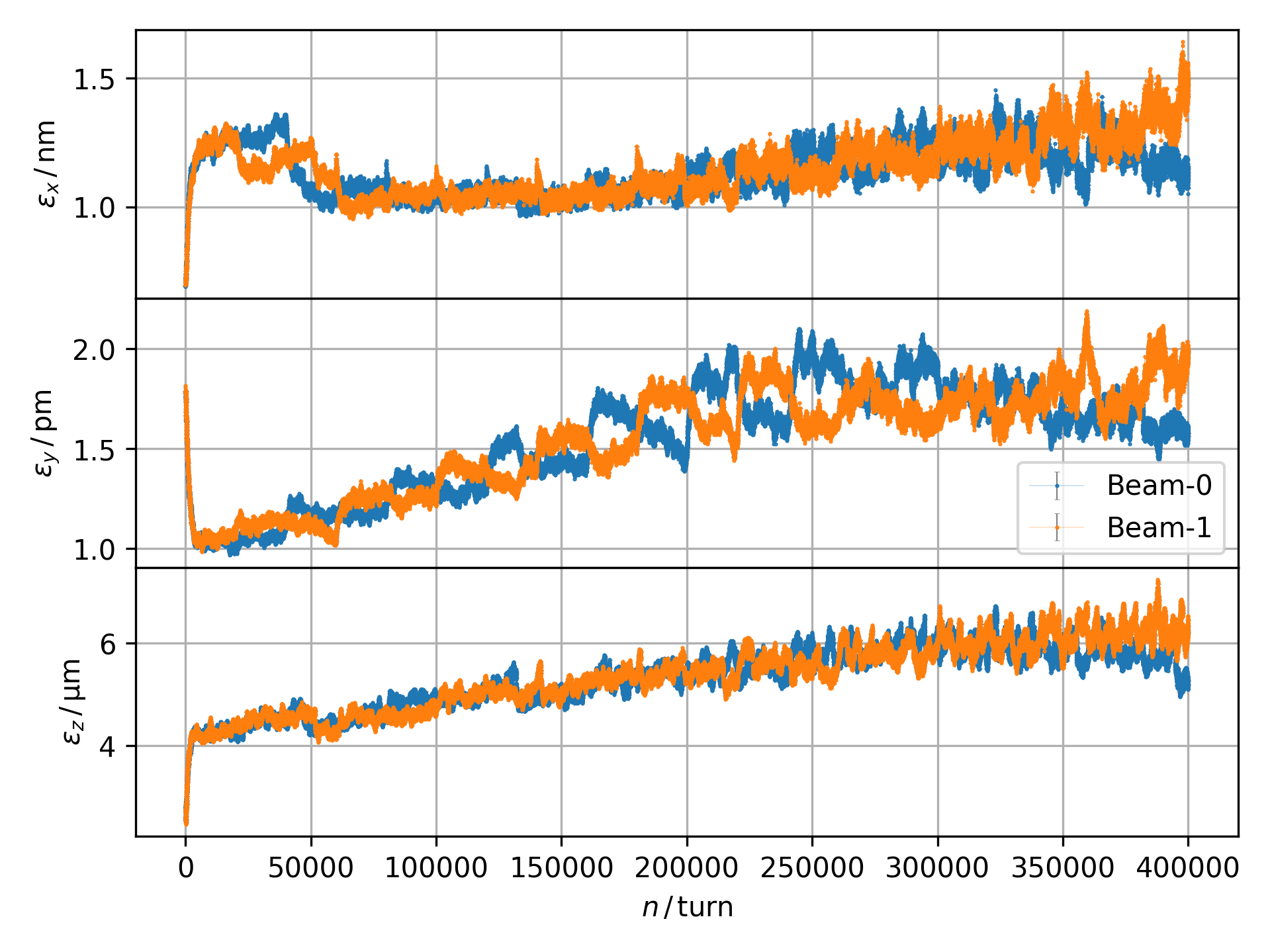}\\
  (\textbf{a})
  \includegraphics[width=\linewidth]{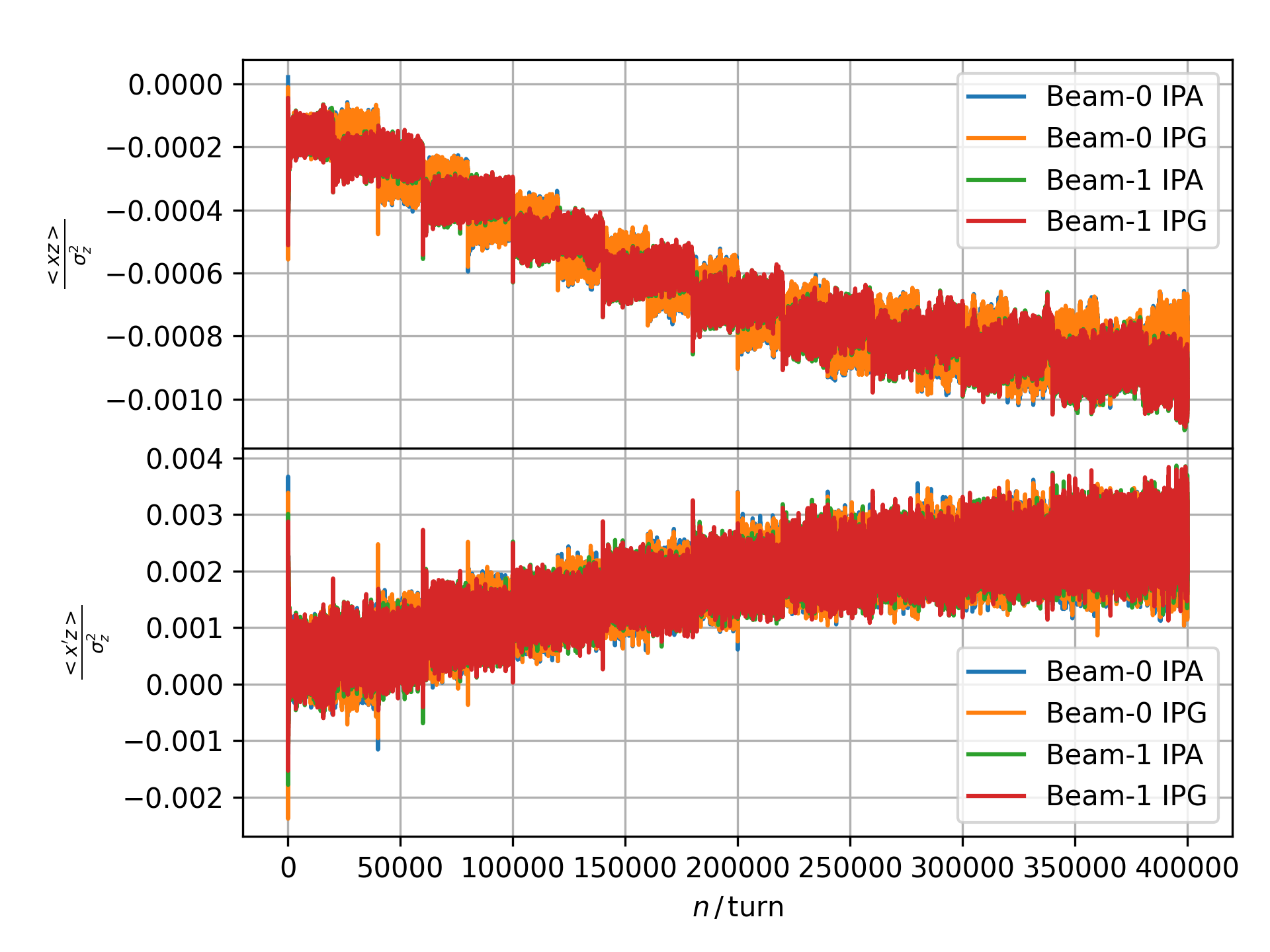}\\
  (\textbf{b})
  \caption{
    Representative indicators of multi-dimensional beam dynamics observed in the Z mode within the quasi-strong-strong simulation.
    (\textbf{a}) Evolution of the transverse and longitudinal emittances during the injection process.
    A pronounced asymmetric behavior among the degrees of freedom is observed as the stored bunch population approaches saturation, exhibiting features characteristic of three-dimensional beam--beam dynamics.
    (\textbf{b}) Time evolution of an $x\text{--}z$ correlation indicator under the same conditions, showing a gradual build-up of transverse--longitudinal coupling over a comparable time scale.
  }
\end{figure}

\subsection{Energy Dependence of Injection Feasibility}
Injection feasibility shows a strong dependence on beam energy, with the Z mode being the most demanding operating point.

The observed behavior is qualitatively consistent with expectations from strong--strong beam--beam interaction,
in which coupled transverse--longitudinal dynamics can lead to enhanced particle loss under weak damping conditions.
Although a direct strong--strong comparison is beyond the scope of this study,
the QSS results provide an approximate description of such effects with significantly reduced computational cost.

\section{Conclusion}
We have presented a study of bootstrapping injection in FCC-ee using a quasi-strong-strong beam--beam scheme implemented in SAD.
Stable injection paths up to the nominal bunch population are identified in the W and H modes, while the Z mode exhibits injection--loss balance within the explored parameter range.
Diagnostic analysis indicates the growing relevance of coupled multi-dimensional dynamics in the low-energy regime.
The present results should therefore be viewed as defining the current boundary of explored parameter space rather than establishing a definitive operational limit.

\bibliographystyle{apsrev4-2}
\bibliography{refs}

\end{document}